\documentclass[aps,prd,onecolumn 10pt,superscriptaddress,floatfix,nofootinbib]{revtex4}
\usepackage{amssymb}
\usepackage{textcomp}
\usepackage{graphicx}
\usepackage{epsfig}
\usepackage{amsmath}
\usepackage{slashed}
\usepackage{units}
\usepackage{hyperref}
\usepackage{array}
\usepackage{lmodern}
\usepackage{natbib}
\usepackage{gensymb}
\usepackage{float}
\usepackage{amsmath}

\begin{document}

\setlength{\parindent}{0pt}

\title{Can neutron disappearance/reappearance experiments definitively rule out the existence of hidden braneworlds endowed with a copy of the Standard Model?}

\author{Coraline Stasser}
\email{coraline.stasser@unamur.be}
\thanks{Corresponding author}
\affiliation{Laboratory for Analysis by Nuclear Reactions, Department of Physics, University
of Namur, 61 rue de Bruxelles, B-5000 Namur, Belgium}

\author{Micha\"{e}l Sarrazin\footnote{Also at Department of Physics, University
of Namur, Belgium}}
\email{michael.sarrazin@ac-besancon.fr}
\thanks{Corresponding author}
\affiliation{Institut UTINAM, CNRS/INSU, UMR 6213, Universit\'{e} de
Bourgogne-Franche-Comt\'{e}, 16 route de Gray, F-25030 Besan\c con Cedex,
France}

\begin{abstract}
Many works, aiming to explain the origin of dark matter or dark energy, consider the existence of hidden (brane)worlds parallel to our own visible world - our usual universe - in a multidimensional bulk. Hidden braneworlds allow for hidden copies of the Standard Model. For instance, atoms hidden in a hidden brane could exist as dark matter candidates. As a way to constrain such hypotheses, the possibility for neutron-hidden neutron swapping can be tested thanks to disappearance-reappearance experiments also known as passing-through-walls neutron experiments. The neutron-hidden neutron coupling $g$ can be constrained from those experiments. While $g$ could be arbitrarily small, previous works involving a $M_4 \times R_1$ bulk, with DGP branes, show that $g$ then possesses a value which is reachable experimentally. It is of crucial interest to know if a reachable value for $g$ is universal or not and to estimate its magnitude. Indeed, it would allow, in a near future, to reject definitively - or not - the existence of hidden braneworlds from experiments. In the present paper, we explore this issue by calculating $g$ for DGP branes, for $M_4 \times S_1/Z_2$, $M_4 \times R_2$ and $M_4 \times T^2$ bulks. As a major result, no disappearance-reappearance experiment would definitively universally rules out the existence of hidden worlds endowed with their own copy of Standard Model particles, excepted for specific scenarios with conditions reachable in future experiments.

\keywords{Brane phenomenology; Hidden braneworld; Neutron-hidden neutron swapping; Neutron disappearance-reappearance.}
\end{abstract}

\maketitle

\section{Introduction}
\label{Intro}

The existence of hidden braneworlds coexisting with our universe in a
multidimensional bulk is an open question often considered in the literature
regarding the quest to explain the dark matter or dark energy conundrum \cite%
{art82, art76, art83, art84, art85, art77, art86, art87}. As a
consequence, beyond cosmological tests or attempts for dark matter particle
detection in astroparticle physics, any other search for direct evidence of
hidden worlds is fundamental. In the last fifteen years, it has been
theoretically shown that neutron swapping could occur between two adjacent
braneworlds both endowed with a copy of the Standard Model of particles \cite%
{art65, art107, art4, art38, art50}. This
phenomenology is related to the fact that any Universe with two braneworlds
-- i.e. two topological defects in the bulk -- is equivalent to an effective
noncommutative two-sheeted spacetime $M_{4}\times Z_{2}$ when one follows
the dynamics of particles below the GeV-scale \cite{art4}. A neutron $n$ can
convert into a hidden neutron $n^{\prime }$ propagating in a hidden
neighboring braneworld with a probability $p\sim g^{2}$, where $g$ is the
coupling constant between the two braneworlds \cite{art38}. As a result, new
kind of experiments exploiting this phenomenon has been suggested in order
to probe the braneworld hypothesis \cite{art40, art41, art5, art6, art24}. For instance,
neutron disappearance (reappearance) toward (from) a hidden brane can be
tested to constrain the coupling constant $g$ between the visible and hidden
sectors. This is the case for instance with passing-through-walls neutron
experiments carried out in the last five years \cite{art5, art6, art24}. Nevertheless, $g$ is a
phenomenological constant which must depend on the brane energy scale (or
its thickness), the interbrane distance, the bulk dimensionality and metrics 
\cite{art50}. As a consequence, knowing the behaviour of $g$ against these
parameters is fundamental to put constraints on specific braneworld
scenarios according to experimental data but also to plan future experiments
and to determine their viability or relevance.

In a previous work \cite{art50}, we introduced a phenomenological approach
to compute the coupling $g$ for two DGP braneworlds \cite{art80, art27} embedded in
a $M_{4}\times R_{1}$ bulk with a warped Chung-Freese-like metric \cite{art29, art104, art105}. One
obtained for the coupling $g$ between neutron and hidden neutron \cite{art50}: 
\begin{equation}
g=(m^{2}/M_{B})\ e^{-md},  \label{g5D}
\end{equation}%
where $m$ is the mass of a constituent quark ($340$ MeV) \cite{art106}, $M_{B}$ the
effective brane energy scale which is related to the thickness of the brane $M_B^{-1}$
with respect to the extra dimension and the ratio of the distortion factors of each brane and $d$ the interbrane distance.

In the present paper, using the same method, we calculate $g$ for various
bulks and we discuss the consequences in regard of experimental data for the
future experiments which can be considered and expected. In section II, we
recall the low-energy framework used to describe a Universe with two
braneworlds at least. In section III, the 5-dimensional case with a $%
S_{1}/Z_{2}$ compactified extra dimension is considered. This scenario has a
historical interest since it is related to the 11D supergravity model of Ho%
\v{r}ava and Witten \cite{art31}. In section IV, the model is extended to 6
dimensions and the expression of $g$ related to two large extra dimensions
is derived. The result is then extended to an ADD-like scenario \cite{art78,
art79} -- thanks to a compactification on a torus $T^{2}$ -- in
section V. Finally, in section VI, magnitudes of the coupling constant $g$
according to these different scenarios are discussed and crossed with experimental data in the context of next generation experiments.

\section{Low-energy description of a Universe with two braneworlds}

\label{framework}

\begin{figure}[tbp]
\centering
\includegraphics[width=8.5 cm]{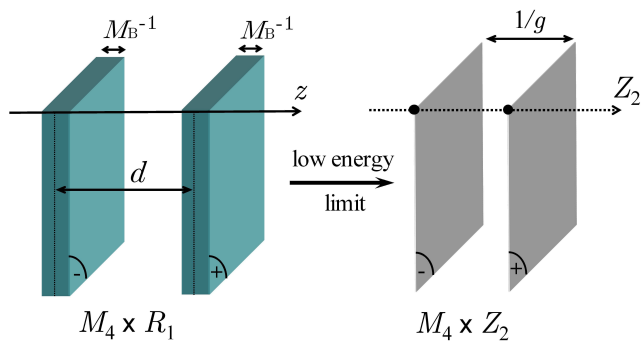}
\caption{(Color online). Sketch of a two-brane universe in a $M_{4}\times
R_{1}$ bulk. Branes are characterized by a thickness $M_{B}^{-1}$ -- where $%
M_{B}$ is the brane energy scale -- with respect to an extra dimension $z$
and an interbrane distance $d$. At low energy, the fermion dynamics in this
universe is the same as in a $M_{4}\times Z_{2}$ non commutative
two-sheeted spacetime where the effective distance $\protect\delta =1/g$ is
related to the coupling constant $g$ between the fermion states localized in
each brane. }
\label{M4Z2}
\end{figure}

The fermion dynamics in a two-braneworld system can be described at low
energy as being the fermion dynamics in a $M_{4}\times Z_{2}$ noncommutative
two-sheeted spacetime as demonstrated elsewhere \cite{art4} (see Fig. \ref
{M4Z2}). The effective two spacetime sheets -- without thickness -- are
separated by an effective distance $\xi =1/g$, where $g$ is the coupling
constant between the fermions of each braneworld. In this $M_{4}\times Z_{2}$ spacetime,
the gauge field $U(1)_{+}\times U(1)_{-}$ arises when considering the
electromagnetic field. Both sheets -- named $(+)$ and $(-)$ -- are endowed
with their own effective gauge field $U(1)_{+}$ and $U(1)_{-}$.
This low energy description is valid whatever the mechanism responsible for
the particle and fields trapping on the branes, the number of
extra dimensions or the metric of the bulk \cite{art4}. The Lagrangian of the $%
M_{4}\times Z_{2}$ model is given by \cite{art4, art38}:

\begin{equation}
\mathcal{L}_{M_{4}\times Z_{2}}\sim \overline{\Psi }\left( {i{\slashed{D}}%
_{A}-M}\right) \Psi ,  \label{L1}
\end{equation}

where

\begin{equation}
{i{\slashed{D}}_{A}-M}=\left( 
\begin{array}{cc}
i\gamma ^{\mu }(\partial _{\mu }+iqA_{\mu }^{+})-m & ig\gamma ^{5}-im_{r} \\ 
ig\gamma ^{5}+im_{r} & i\gamma ^{\mu }(\partial _{\mu }+iqA_{\mu }^{-})-m%
\end{array}%
\right)  \label{Dirac}
\end{equation}%
is typical of the noncommutative $M_{4}\times Z_{2}$ spacetime. For these
two equations, $\Psi =\left( 
\begin{array}{c}
\psi _{+} \\ 
\psi _{-}%
\end{array}%
\right) $ is a two-level spinor which contains the fermionic wave functions $%
\psi _{+}$ in the visible brane ($+$) and $\psi _{-}$ in the hidden brane ($-$). 
$A_{\mu }^{\pm }$ are the electromagnetic four-potentials on each brane $%
(\pm )$ resulting from the gauge field $U(1)_{+}\times U(1)_{-}$. $m$ is the
mass of the bounded fermion on a brane. $m_{r}$ is a mass-mixing term whose
the phenomenology can be neglected when compared with the one induced by the coupling
constant $g$ as shown in previous works \cite{art4, art38}. This last coupling induces a mixing
which leads to fast Rabi oscillations between fermions of each braneworld,
with a probability $p\sim g^{2}$ \cite{art38}. Most important, $g$ can be
calculated from the fondamental properties of the two-braneworld Universe as
mentioned above, i.e. $g$ depends on the brane energy scale and on the
interbrane distance in the bulk \cite{art4, art50}, for instance. The derivation of $g$ against
these parameters is considered in the following for given bulks of interest.

\section{Neutron-hidden neutron coupling in a $M_4 \times S_1/Z_2$ bulk}

\label{5D}

\begin{figure}[ht]
\centering
\includegraphics[width=8.5 cm]{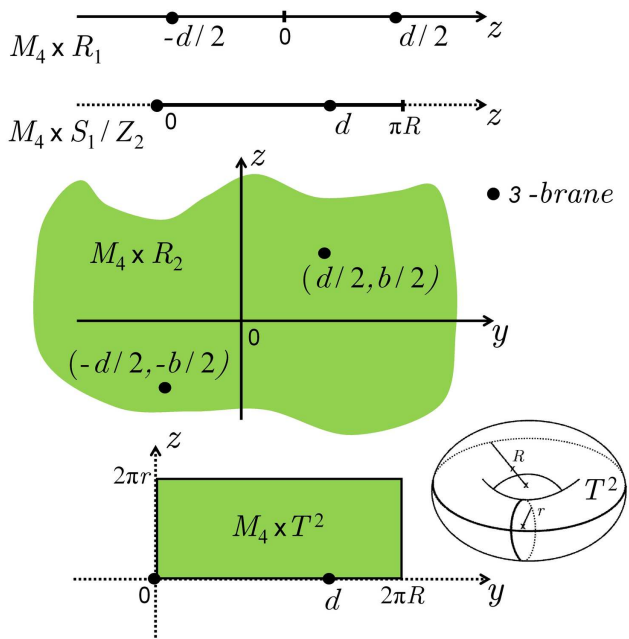}
\caption{(Color online). Sketches of the extra-dimensional parts of the
bulks under consideration for a two-braneworld Universe.}
\label{Fig2}
\end{figure}

This section pursues the phenomenological investigations, regarding a
two-braneworld Universe in a 5-dimensional bulk, introduced in our
preliminary work \cite{art50} and in which the coupling constant $g$ for a
neutron was computed for a $SO(3,1)$-broken 5-dimensional $M_{4}\times R_{1}$
bulk. In this paper, a similar calculation is derived, but now for a 5D $%
M_{4}\times S_{1}/Z_{2}$ orbifold bulk. The interest of such a scenario arises
from the supergravity model of Ho\v{r}ava-Witten \cite{art31}. Their
approach makes possible the link between the $E_{8}\times E_{8}$ heterotic
super-string theory in 10 dimensions and the 11-dimensional supergravity on
the orbifold $M_{10}\times S_{1}/Z_{2}$, where 6 of the 11 dimensions are
compactified on a Calabi-Yau manifold. At low energy, this model leads then
to a $M_{4}\times S_{1}/Z_{2}$ Universe with two 3-branes localized at the
boundaries of the $S_{1}/Z_{2}$ orbifold. Such a configuration is considered
for instance in ekpyrotic scenarios \cite{art60, art73}, in the Randall-Sundrum I model \cite{art88}, or in
the Chung-Freese approach \cite{art29}. In these models, a warped metric is often
included. Nevertheless, as shown in our previous work \cite{art50}, while
there is a bare brane energy scale (or bare brane thickness) for a flat
metric, the warped metric induces an effective brane energy scale (or
effective brane thickness). Experimentally, bare and effective energy scales
cannot be distinguish from each other. As a consequence, in the present
work, all calculations are made with a Minkowski metric.

The phenomenological model here under consideration, and related
calculations, are fully introduced and described elsewhere \cite{art50} for
a $M_{4}\times R_{1}$ bulk. Nevertheless, the reader will find more details
in section \ref{6D} since they are necessary to explain how to deal with 6-dimensional bulks. As
basic hypotheses, one considers fermion sectors $\psi _{+}$ and $\psi _{-}$
respectively which exist only on branes $(+)$ and $(-)$ respectively, and a
massless fermion sector $\Psi $ able to propagate through the whole bulk.
Each sectors are coupled to each other on each brane through the action \cite%
{art50}: 
\begin{eqnarray}
\label{Sc}
S_{coupling} &=&-\int d^{4}xdz\sqrt{\left\vert g^{(4)}\right\vert }
 \\
&&\times \left\{ \frac{m}{M_{B}^{1/2}}\left( \overline{\psi }_{+}\Psi +%
\overline{\Psi }\psi _{+}\right) \delta (z-d/2)\right.   \notag \\
&&\left. +\frac{m}{M_{B}^{1/2}}\left( \overline{\psi }_{-}\Psi +\overline{%
\Psi }\psi _{-}\right) \delta (z+d/2)\right\} ,  \notag
\end{eqnarray}%
for two branes located at $z=\pm d/2$ for instance (see Fig. \ref{Fig2}). Let us call $%
G(z|z^{\prime })$ the propagator of the bulk sector $\Psi $ along the extra
dimension. The bulk Dirac matrices $\Gamma ^{A}$ are such that $\left\{
\Gamma ^{A},\Gamma ^{B}\right\} =2\eta ^{AB}\mathbf{1}_{4\times 4}$ ($%
A,B=0,1,\ldots ,4$) with $\eta ^{AB}$ the Minkowski metric with a $%
(+,-,-,-,-)$ signature, and $\Gamma ^{4}=-i\gamma _{5}$. Then, it can be
proved \cite{art50} that the coupling constant $g$ is equal to the component
of $2(m^{2}/M_{B})G(d/2|-d/2)$ proportional to $\gamma _{5}$. For instance,
Eq. (\ref{g5D}) is obtained by considering the bulk sector propagator along
the extra dimension $R_{1}$ in a $M_{4}\times R_{1}$ bulk \cite{art50}: 
\begin{eqnarray}
G(z) &=&\frac{1}{2\pi }\int \frac{i\gamma ^{5}\kappa +m}{\kappa ^{2}+m^{2}}%
e^{-i\kappa z}d\kappa   \notag \\
&=&(1/2)e^{-m\left\vert z\right\vert }\left( \mathbf{1}+sign(z)\gamma
^{5}\right) ,  \label{greenfuncprevious}
\end{eqnarray}%
For a $M_{4}\times S_{1}/Z_{2}$ bulk, the propagator expression given by Eq.
(\ref{greenfuncprevious}) is no longer valid. The $S_{1}/Z_{2}$ symmetry
must be taken into account. First, the $G_{S_{1}}(z)$ propagator, along $%
S_{1}$ only, can be easily obtained from $G(z)$ thanks to a periodic
summation with period $2\pi R$ \cite{art90}:

\begin{equation}
G_{S_{1}}(z)=\displaystyle\sum_{n=-\infty }^{+\infty }G(z+n\ 2\pi R).
\label{propaS1}
\end{equation}%
Then, the $G_{S_{1}/Z_{2}}(z)$ propagator can be found thanks to the
relationship linking the propagator to the bulk field eigenstates \cite{art90}:

\begin{equation}
G_{S_{1}/Z_{2}}(z)=\langle \Psi (z)\bar{\Psi}(z)\rangle ,
\label{propaeigenstate}
\end{equation}%
where $\Psi (z)$ is a linear combination of the two possible solutions
induced by a $Z_{2}$ symmetry:

\begin{equation}
\Psi (z)=\frac{1}{2}(\chi (z)+\gamma ^{5}\chi (-z)),  \label{Z2sym}
\end{equation}%
with $\chi $ the periodical eigenstate with period $2\pi R$ resulting of the $%
S_{1}$ symmetry. From Eqs. (\ref{propaeigenstate}) and (\ref{Z2sym}) it is possible
to deduce the $S_{1}/Z_{2}$ propagator from $G_{S_{1}}$, and one obtains:

\begin{eqnarray}
G_{S_{1}/Z_{2}}(z|z^{\prime }) &=&\frac{1}{4}\Bigg\{G_{S_{1}}(z-z^{\prime
})-G_{S_{1}}(-z+z^{\prime }) \notag\\
&& -G_{S_{1}}(z+z^{\prime })\gamma ^{5} + G_{S_{1}}(-z-z^{\prime })\gamma ^{5}%
\Bigg\}.  \label{gfunction1}
\end{eqnarray}

From Eqs. (\ref{greenfuncprevious}), (\ref{propaS1}) and (\ref{gfunction1}), it results
the following expression:

\begin{eqnarray}
G_{S_{1}/Z_{2}}(z|z^{\prime }) &=&\frac{1}{4}\displaystyle\sum_{n=-\infty
}^{+\infty }\Bigg\{e^{-m\left\vert z-z^{\prime }-2\pi Rn\right\vert
}sign(z-z^{\prime }-2\pi Rn)\gamma ^{5}  \notag \\
&&-e^{-m\left\vert z+z^{\prime }-2\pi Rn\right\vert }sign(z+z^{\prime }-2\pi
Rn)\Bigg\},  \label{gfunction2}
\end{eqnarray}%
with $\Gamma ^{0}G^{\dagger }(z)\Gamma ^{0}=G(-z)$. Following the same
procedure than in our previous paper \cite{art50} by using the propagator
expressed by Eq. (\ref{gfunction2}), the coupling constant $g$ can be
evaluated against the position of the braneworlds on the orbifold $%
S_{1}/Z_{2}$ (see Fig. \ref{Fig2}). When the branes are localized at the orbifold limits, i.e. at $%
z=0$ and $z=\pi R$, the coupling constant $g$ drops to $0$, meaning that no
geometrical coupling is allowed in such a situation. But if one considers our
brane located at $z=0$ and the hidden one at $z=d$ (where $d\in \left] 0,\pi
R\right[ $) -- i.e. the hidden brane lurks along $S_{1}/Z_{2}$ -- the
coupling constant between neutron and hidden neutron is now given by:

\begin{equation}
g=\frac{m^{2}}{M_{B}}\left( \frac{e^{md}-e^{-md+m2\pi R}}{1-e^{m2\pi R}}%
\right) ,  \label{g1}
\end{equation}%
with $m$ the mass of the quark constituent ($340$ MeV \cite{art106}), $M_{B}$ the brane
energy scale and $R$
the compactification radius. For $R\rightarrow +\infty $, we retrieve the
non-compactified 5-dimensional case given by expression \ref{g5D} and for $%
d\rightarrow \pi R$, we retrieve $g=0$.

\section{Neutron-hidden neutron coupling in a flat non-compact 6D bulk}

\label{6D}

\bigskip As a beginning and a prerequisite, let us now describe the
coupling between two braneworlds in a flat non-compact 6-dimensional bulk.
We follow the same approach as previously \cite{art50}. This will allow us
to consider in section \ref{ADD} the coupling between each brane in an ADD-like scenario 
\cite{art78, art79}. We consider two 3-branes respectively located at 
$(y,z)=(d/2,b/2)$ and $(y,z)=(-d/2,-b/2)$ (see Fig. \ref{Fig2}). The coupling action $S_{c}$
between the 6D brane sectors $\Psi _{\pm }$ and the 6D bulk sector $\Psi $
is now given by: 
\begin{eqnarray}
S_{c} &=&\int d^{6}x\Bigg\{-\frac{m}{M_{B}}(\overline{\Psi }_{+}\Psi +%
\overline{\Psi }\Psi _{+})\delta (y-d/2)\delta (z-b/2)  \label{6Dc} \\
&&-\frac{m}{M_{B}}(\overline{\Psi }_{-}\Psi +\overline{\Psi }\Psi
_{-})\delta (y+d/2)\delta (z+b/2)\Bigg\}.  \notag
\end{eqnarray}%
The energy scale of the branes $M_{B}$, with $M_B^{-1}$ the
extradimensional extent of the braneworld in the bulk, is introduced to take
into account the extradimensional volume in which the coupling interaction occurs. By contrast with Eq. (\ref{Sc}), the power $1$ of $M_{B}$
ensures the correct dimensionality of the problem. The braneworld $S_{\pm }$
action is given by:

\begin{equation}
S_{\pm }=\int d^{4}x\ \overline{\psi }_{\pm }\ (i\gamma ^{\mu }\partial
_{\mu }+iqA_{\mu }^{\pm }-m)\ \psi _{\pm },  \label{brane4D}
\end{equation}%
where $\mu =0,1,2,3$ and with $\psi _{\pm }$ the 4D brane sectors and $%
A^{\pm }$ the electromagnetic vector potentials on each brane. The bulk
field $\Psi $ action $S_{bulk}$ is given by: 
\begin{equation}
S_{bulk}=\int d^{6}x\ \overline{\Psi }\left( i\Gamma ^{A}\left( \partial
_{A}+iq\mathcal{A}_{A}\right) \right) \Psi ,  \label{6Daction}
\end{equation}%
where $A=0,1,2,3,5,6$ and $\mathcal{A}_{A}$ is the electromagnetic vector potential of
the bulk. It is noteworthy that the electromagnetic potential is assumed to
exist only on the braneworlds, with $\mathcal{A}_{4}=\mathcal{A}_{5}=0$ \cite{art4}. Here the bulk Dirac
matrices $\Gamma ^{A}$ are such that $\left\{ \Gamma ^{A},\Gamma
^{B}\right\} =2\eta ^{AB}\mathbf{1}_{8\times 8}$ with $\eta ^{AB}$ the
Minkowski metric with a $(+,-,-,-,-,-)$ signature. We then use the following
8-dimensional Dirac matrices $\Gamma ^{A}$: 
\begin{eqnarray}
\Gamma ^{\mu } &=&\sigma _{1}\otimes \gamma ^{\mu }=\left( 
\begin{array}{cc}
0 & \gamma ^{\mu } \\ 
\gamma ^{\mu } & 0%
\end{array}%
\right) ; \\
\text{ }\Gamma ^{5} &=&-i\sigma _{1}\otimes \gamma ^{5}=\left( 
\begin{array}{cc}
0 & -i\gamma ^{5} \\ 
-i\gamma ^{5} & 0%
\end{array}%
\right) ;\text{ } \\
\Gamma ^{6} &=&-i\sigma _{2}\otimes \mathbf{1}_{4\times 4}=\left( 
\begin{array}{cc}
0 & -\mathbf{1}_{4\times 4} \\ 
\mathbf{1}_{4\times 4} & 0%
\end{array}%
\right) ,
\end{eqnarray}%
and such that the chiral matrice $\Gamma ^{7}=-\Gamma ^{0}\Gamma ^{1}\Gamma
^{2}\Gamma ^{3}\Gamma ^{5}\Gamma ^{6}$ is given by: 
\begin{equation}
\Gamma ^{7}=\sigma _{3}\otimes \mathbf{1}_{4\times 4}=\left( 
\begin{array}{cc}
\mathbf{1}_{4\times 4} & 0 \\ 
0 & -\mathbf{1}_{4\times 4}%
\end{array}%
\right) .
\end{equation}%
The chiral 6D states are then defined through: 
\begin{equation}
\left( \frac{1\pm \Gamma ^{7}}{2}\right) \Psi =\Psi _{R/L},
\end{equation}%
and $\Psi $ can be written as: 
\begin{equation}
\Psi =\left( 
\begin{array}{c}
\psi _{R} \\ 
\psi _{L}%
\end{array}%
\right) \mathrm{such\  that} \   \Psi _{R}=\left( 
\begin{array}{c}
\psi _{R} \\ 
0%
\end{array}%
\right) \mathrm{and} \   \Psi _{L}=\left( 
\begin{array}{c}
0 \\ 
\psi _{L}%
\end{array}%
\right) \mathrm{.}
\end{equation}%

Now, regarding the 6D brane sectors $\Psi _{\pm }$, the 6D chiral states
do not correspond to the 4-dimensional ones on the branes. Indeed, the 6D
chirality appears as an extra quantum number added to the usual particles of
the Standard Model on a brane, thus doubling the Standard Model on the
brane, a situation which is not observed. As a consequence, we assume that
each brane can only support one 6D chiral state, for instance the left one,
while the other cannot be trapped on the brane. Such a situation is
supported by the works dealing with the domain wall description of branes
where it is known that the fermions' trapping on branes depends on their
chirality \cite{art26, art91, art92, art93, art94, art95, art96, art97, art98, art99, art100, art101, art4, art102, art103}. As an ansatz, the brane sectors are given by:\bigskip 
\begin{equation}
\Psi _{\pm }=\left( 
\begin{array}{c}
0 \\ 
\psi _{\pm }%
\end{array}%
\right) ,
\end{equation}%
and where $\psi _{\pm }$ follow the action given by Eq. (\ref{brane4D})
thanks to the above choice for the gamma matrices and in accordance with the
6D action in Eq. (\ref{6Daction}).

\bigskip Now, from the whole action, the bulk field follows:

\begin{equation}
\left( i\Gamma ^{A}\left( \partial _{A}+iq\mathcal{A}_A\right) \right) \Psi =\frac{m%
}{M_{B}}\Psi _{+}\delta (y-d/2)\delta (z-b/2)+\frac{m}{M_{B}}\Psi _{-}\delta
(y+d/2)\delta (z+b/2),  \label{eqbulk6D}
\end{equation}%
where the fields on each braneworld act as sources (or wells) for the bulk field.

From Eq. (\ref{eqbulk6D}) and using the mass shell condition \cite{art50} $\left( i\Gamma _{\pm }^{\mu
}\left( \partial _{\mu }+iqA_{\mu }\right) -m\right) \Psi =0$, one deduces
the following propagator for the bulk sector along extra dimensions:

\begin{eqnarray}
G(y,z) &=&\frac{1}{4\pi ^{2}}\int \int \frac{\Gamma ^{5}q+\Gamma ^{6}\kappa
+m}{q^{2}+\kappa ^{2}+m^{2}}e^{i\kappa z}e^{iqy}d\kappa dq  \label{6Dprop} \\
&=&\frac{1}{2\pi }\frac{im}{\sqrt{z^{2}+y^{2}}}K_{1}(m\sqrt{z^{2}+y^{2}}%
)\left( sign(y)\ |y|\ \Gamma ^{5}+sign(z)\ |z|\ \Gamma ^{6}\right)  \notag \\
&&+\frac{m}{2\pi }K_{0}(m\sqrt{z^{2}+y^{2}}),  \notag
\end{eqnarray}%
with $\Gamma ^{0}G^{\dagger }(y,z)\Gamma ^{0}=G(-y,-z)$ and from which $\Psi 
$ can be expressed thanks to Eq. (\ref{eqbulk6D}):%
\begin{equation}
\Psi (x,y,z)=\frac{m}{M_{B}}G(y-d/2,z-b/2)\Psi _{+}(x)+\frac{m}{M_{B}}%
G(y+d/2,z+b/2)\Psi _{-}(x).  \label{Psibulk6D}
\end{equation}%
Injecting Eq. (\ref{Psibulk6D}) in the coupling action given by Eq. (\ref{6Dc}) and
looking for the $M_{4}\times Z_{2}$ effective action $S_{M_{4}\times
Z_{2}}=S_{+}+S_{-}+S_{c}$ given by Eqs. (\ref{L1}) and (\ref{Dirac}), one
successively gets: 
\begin{equation}
S_{c}=-\frac{2m^{2}}{M_{B}^{2}}\int d^{4}x\Bigg\{\overline{\Psi }%
_{+}G(d,b)\Psi _{-}+\overline{\Psi }_{-}G(-d,-b)\Psi _{+}\Bigg\}
\label{Scp}
\end{equation}%
and%
\begin{eqnarray}
S_{M_{4}\times Z_{2}} &=&\int d^{4}x\ \Bigg\{ \overline{\psi }_{+}\ (i\gamma ^{\mu
}\partial _{\mu }+iqA_{\mu }^{+}-m)\ \psi _{+}  \label{SM4Z26D} \\
&&+\overline{\psi }_{-}\ (i\gamma ^{\mu }\partial _{\mu }+iqA_{\mu }^{-}-m)\
\psi _{-}  \notag \\
&&+ig\overline{\psi }_{+}\gamma ^{5}\psi _{-}+ig\overline{\psi }_{-}\gamma
^{5}\psi _{+}  \notag \\
&&-im_{r}\overline{\psi }_{+}\psi _{-}+im_{r}\overline{\psi }_{-}\psi _{+} \Bigg\}, 
\notag
\end{eqnarray}%
with the coupling constant $g$ given by: 
\begin{equation}
g=\frac{m^{3}}{\pi M_{B}^{2}}\frac{d}{D}K_{1}(mD),  \label{6gDgen}
\end{equation}%
and 
\begin{equation}
m_{r}=\frac{m^{3}}{\pi M_{B}^{2}}\frac{b}{D}K_{1}(mD),  \label{mr}
\end{equation}%
where $D$ is the distance between the two braneworlds (see Fig. \ref{Fig2}), with $%
D=\sqrt{d^{2}+b^{2}}$ and $m$ the mass of a constituent quark ($340$ MeV) when considering the neutron-hidden neutron coupling \cite{art106,art50}. It is interesting to note that due to
the bulk symmetry breaking induced by the branes regarding to the 6D
chirality, $g$ or $m_{r}$ can vanishe depending on the value of $d$ or $b$.
For instance, if $b=0$, $m_{r}$ is now equal to zero, and Eq. (\ref{6gDgen})
reduces to: 
\begin{equation}
g=\frac{m^{3}}{\pi M_{B}^{2}}K_{1}(md).  \label{g6D}
\end{equation}

\section{Neutron-hidden neutron coupling in an ADD bulk}
\label{ADD}

The last case introduced in this paper is the compactification of the
two extra dimensions on a torus ($T^{2}\equiv S_{1}\times S_{1}$
manifold), with two 3-branes respectively located at $(y,z)=(0,0)$ and $%
(y,z)=(d,0)$, with $d\in \ ]0,2 \pi R[$ (see Fig. \ref{Fig2}). This model is of interest as it is
reminiscent of the ADD scenario \cite{art78, art79} but with two
branes. Using the same approach as in section \ref{5D} to derive the
propagator in a compactified bulk, the propagator for the bulk sector on the
torus can be deduced from Eq. (\ref{6Dprop}), and we get: 
\begin{eqnarray}
&&G(y,z)=\displaystyle\sum_{n=-\infty }^{+\infty }\displaystyle%
\sum_{k=-\infty }^{+\infty }\Big\{\frac{1}{2\pi }\frac{imK_{1}(m\sqrt{%
(y+n2\pi R)^{2}+(z+k2\pi r)^{2}})}{\sqrt{(y+n2\pi R)^{2}+(z+k2\pi r)^{2}}} \\
&&\times \left( sign(y+n2\pi R)\ |y+n2\pi R|\ \Gamma ^{5}+sign(z+k2\pi r)\
|z+k2\pi r|\ \Gamma ^{6}\right) \notag  \\
&&+\frac{m}{2\pi }K_{0}(m\sqrt{(y+n2\pi R)^{2}+(z+k2\pi r)^{2}})\Big\} \notag ,
\end{eqnarray}%
with $r$ and $R$ the compactification radii of the extra dimensions (see Fig. \ref{Fig2}). It leads to the following coupling constant $g$ expression:%
\begin{eqnarray}
g &=&\frac{m^{3}}{\pi M_{B}^{2}}\displaystyle\sum_{n=-\infty }^{+\infty }%
\displaystyle\sum_{k=-\infty }^{+\infty }\frac{K_{1}(m\sqrt{(d+n2\pi
R)^{2}+(k2\pi r)^{2}})}{\sqrt{(d+n2\pi R)^{2}+(k2\pi r)^{2}}}  \\
&&\times sign(d+n2\pi R)\ |d+n2\pi R|\  \notag  \label{gtore}
\end{eqnarray}%
for which there is no trivial expression. It is notheworthy that, as for the
compact 5-dimensional case introduced in section \ref{5D}, some locations ($d=\pi R$ for instance) of
the braneworlds cancel the coupling. When $r,R\rightarrow +\infty $, i.e.
the torus tends towards a plane, all terms in the summation in Eq. (\ref%
{gtore}) tend towards zero, except for $(n,k)=(0,0)$, thus leading to the
expected expression Eq. (\ref{g6D}).

\section{Discussion}

\label{coupling}

\begin{figure} [h!]
\begin{center}
\includegraphics[scale=0.46]{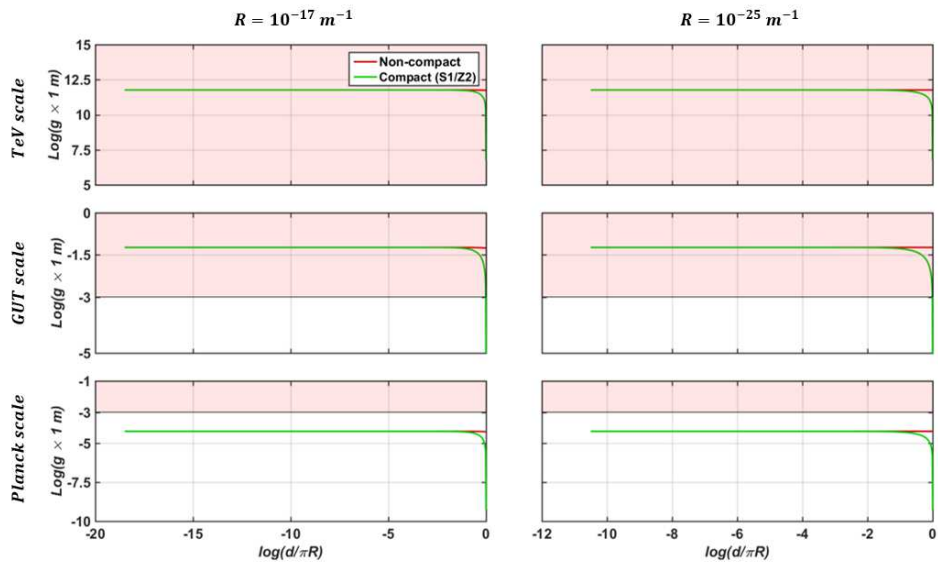}
\end{center}
\caption{(Color online). Neutron-hidden neutron coupling constant $g$ against interbrane
distance $d\in \ ]0,\pi R[$ for a 5-dimensional bulk with 1 extra dimension compactified on
a $S_1/Z_2$ orbifold. $g$ is plotted for three braneworld energy
scales (the TeV scale, the GUT scale and the Planck scale) and for two
compactification radii ($R$) chosen arbitrarily ($R=10^{-17}$ m$%
^{-1}$ and $R=10^{-25}$ m$^{-1}$). Red curves represent the non-compact 5-dimensional case and green curves the $%
S_1/Z_2$ compact case. Red regions for values greater than $g=200$ peV (or $%
10^{-3}$ m$^{-1}$ in natural units) are excluded with confidence from
experimental data \cite{art6}. For interbrane distances greater than 
$0.5$ fm, neutron exchange is supposed to be precluded ($g=0$ m$^{-1}$) by
the model \cite{art50}.}
\label{Fig22}
\end{figure}

\begin{figure} [h!]
\centering
\includegraphics[scale=0.45]{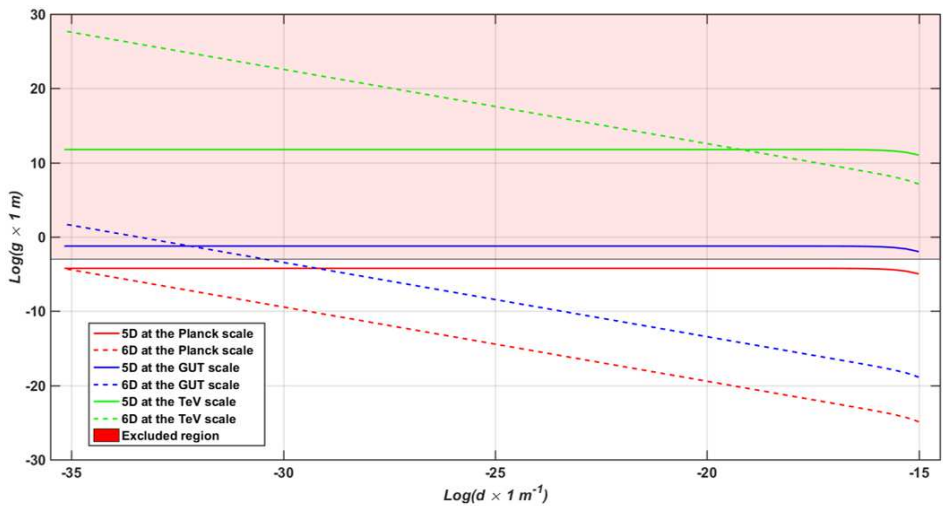}
\caption{(Color online). Neutron-hidden neutron coupling constant $g$ against
interbrane distance $d$ for non-compact 5-dimensional and 6-dimensional bulks and
for various brane energy scales $M_B$, i.e the TeV scale, the GUT scale
and the Planck scale. Solid lines represent the 5-dimensional bulk and dash
curves the 6-dimensional bulk. Red regions for values greater than $g=200$
peV (or $10^{-3}$ m$^{-1}$ in natural units) are ruled out from stringent experimental data \cite{art6}. For interbrane distances greater
than $0.5$ fm, neutron exchange is supposed to be precluded ($g=0$ m$^{-1}$)
by the model \cite{art50}.}
\label{Fig1}
\end{figure}

\begin{figure} [h!]
\includegraphics[scale=0.47]{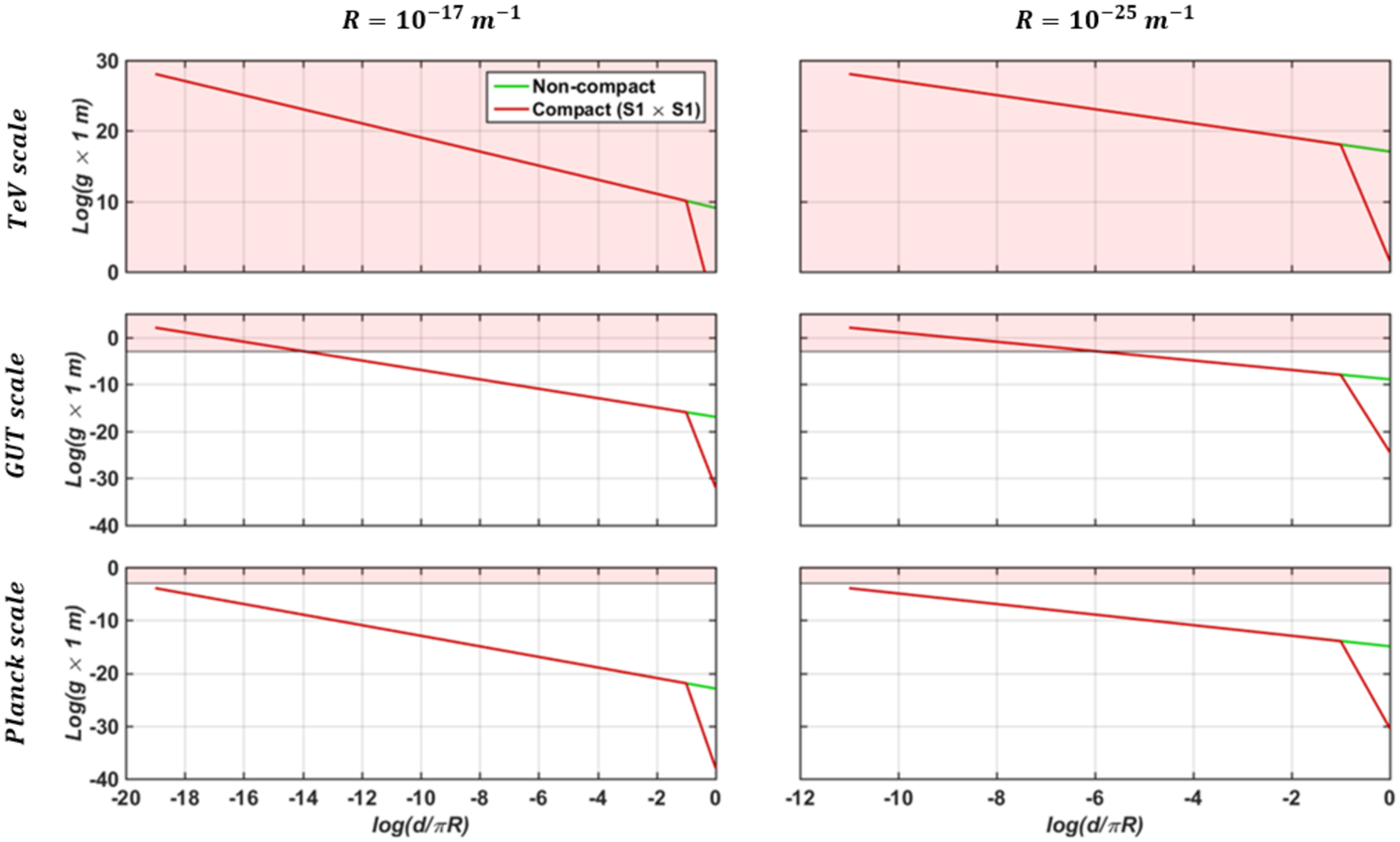}
\caption{(Color online). Neutron-hidden neutron coupling constant $g$ against interbrane
distance  $d\in \ ]0,\pi R[$ for a 6-dimensional bulk with 2 extra dimensions compactified on
a $S_1 \times S_1$ manifold, i.e. a torus. $g$ is plotted for three braneworld energy
scales (the TeV scale, the GUT scale and the Planck scale) and for two
compactification radii ($r$ and $R$) chosen arbitrarily ($R=r=10^{-17}$ m$%
^{-1}$ and $R=r=10^{-25}$ m$^{-1}$). Red regions for
values greater than $g=200$ peV (or $10^{-3}$ m$^{-1}$ in natural units) are
excluded with confidence from experimental data \protect\cite{art6}. For
interbrane distances greater than $0.5$ fm, neutron exchange is supposed to
be precluded ($g=0$ m$^{-1}$) by the model \protect\cite{art50}.}
\label{Fig3}
\end{figure}

The disappearance of a geometrical coupling for two braneworlds located at
the $S_1/Z_2$ orbifold limits makes impossible to constrain the Ho\v{r}%
ava-Witten 11-dimensional supergravity \cite{art31} with neutron-hidden
neutron transitions. However, for any other locations of the branes, the
expression of the coupling constant is given by Eq. (\ref{g1}). Such locations
are allowed in the context of some ekpyrotic scenarios \cite{art60, art73}. Figure \ref{Fig22}
shows the behavior of the neutron coupling constant against the
interbrane distance $d$ for an extra dimension compactified on a $S_1/Z_2$
orbifold (green points derived from Eq. (\ref{g1}), compared to the non-compact case 
\cite{art50} (red line). In Fig. \ref{Fig22}, $g$ is plotted for two
compactification radii (chosen arbitrarily), i.e $R=10^{-17}$ m$^{-1}$ and $%
R=10^{-25}$ m$^{-1}$, and for three brane energy scales: the TeV scale, the
GUT scale and the Planck scale. For the TeV scale, the $S_1/Z_2$ compact
case is ruled out as well as the non-compact case. For the GUT scale, the drop of the coupling for the compact case makes impossible to exclude this scenario for
interbrane distances $d \rightarrow \pi R$. For
the Planck energy scale, the non-compact case is very close to be
excluded with future passing-through-walls neutron experiments \cite{art6, art24} while significant improvements are needed to rule out the compact case for interbrane distances close to $\pi R$ .

Figure \ref{Fig1} shows the neutron-hidden neutron coupling constant $g$ in function of the
interbrane distance in a non-compact bulk for one extra dimension (from Eq. (\ref%
{g5D})) and for two extra dimensions (from Eq. (\ref{g6D})). Here again, three
braneworld energy scales $M_B$ are also considered: the TeV scale, the GUT scale
and the Planck scale. Braneworlds related to the TeV energy
scale are fully excluded for one as well as for two extra dimensions.
Braneworlds related to the GUT energy scale are also ruled out for one
extra dimension. As shown in Fig. \ref{Fig1}, the transition from a
5-dimensional bulk to a 6-dimensional one significantly reduces the coupling constant values for GUT and Planck scales. While the Planck scale
for one non-compact extra dimension is almost reachable by experiments \cite{art6}, the 6-dimensional case is far beyond the sensitivity of passing-through-walls neutron experiments \cite{art6}. The present results show the impossibility
for current passing-through-walls neutron experiments to constrain all the range of
interbrane distances for GUT and Planck scales for bulks with more than 5 dimensions.
Indeed, the swapping probability $p$ (see sections \ref{Intro} and \ref{framework}) and the coupling constant $g$ are
related as $g=\sqrt{p}$. While a gain of a factor $10$ on the last constrain
found in 2016 ($p<4.6 \times 10^{-10}$ at 95\% CL) is expected for future passing-through-walls neutron experiments, the 6-dimensional case is far to be
reachable by such experiments.

Finally, Fig. \ref{Fig3} shows the coupling constant $g$ against the
interbrane distance $d$ for two extra dimensions compactified on a $S_1
\times S_1$ manifold, i.e. a torus. As previously, we explore the same three energy scales (TeV, GUT and Planck scales). Two compactification radii are
chosen (arbitrarily): $R=r=10^{-17}$ m$^{-1}$ and $R=r=10^{-25}$ m$^{-1}$.
As shown by Fig. \ref{Fig3}, the compactification leads to a decrease of the coupling for
interbrane distances  $d \rightarrow \pi R$ with respect to the non-compact case. While the TeV energy scale is completely ruled out whatever the values
of compactification radii, all the parameter range of GUT and Planck scales
are unreachable for the sensitivity of current and future passing-through-walls neutron experiments.

\section{Conclusions}

Many scenarios consider hidden braneworlds in the vicinity of our visible one, living together in a N-dimensional bulk. Here we have described the behavior of the neutron-hidden neutron coupling constant – an experimentally measurable parameter – for various bulks. It has been first shown that the Ho%
\v{r}ava-Witten 11-dimensional supergravity and related models cannot be excluded with passing-through-walls neutron experiments. But it is not the case for some ekpyrotic scenarios provided that one braneworld, at least, is not located on a boundary of the $M_4 \times S_1/Z_2$ orbifold. Next, we have considered 6D bulks, with two extralarge extra dimensions or compactified on a torus in an ADD-like configuration. The addition of more than one extra dimension significantly drops the coupling constant values, making possible yet to test these scenarios but precluding to fully rule out the whole range of braneworld models. While braneworlds endowed with their own copy of the Standard Model at a TeV energy scale are already experimentally excluded, those at GUT or Planck scale are still reachable and their existence could be either confirmed or rejected for 5D bulks. By contrast, future experiments involving neutron disappearance-reappearance could constrain 6D bulks scenarios but cannot totally exclude them. Such a situation prevents to definitively close these lines of theoretical research.

\section*{Acknowledgements}

C.S. is supported by a FRIA doctoral grant from the Belgian F.R.S-FNRS.

\bibliographystyle{unsrt}
\bibliography{Biblio}

\end{document}